\documentclass[aps,twocolumn,pra,showpacs,floatfix]{revtex4}

\usepackage[dvips]{graphicx}

\begin{document}


\title{Frequency shift of hyperfine transitions due to blackbody radiation}

\author{E. J. Angstmann}
\affiliation{School of Physics, University of New South Wales,
Sydney 2052, Australia}
\author{V. A. Dzuba}
\affiliation{School of Physics, University of New South Wales,
Sydney 2052, Australia}
\author{V. V. Flambaum}
\affiliation{School of Physics, University of New South Wales,
Sydney 2052, Australia}

\date{\today}

\begin{abstract}
We have performed calculations of the size of the frequency shift
induced by a static electric field on the clock transition
frequencies of the hyperfine splitting in Yb$^{+}$, Rb, Cs,
Ba$^{+}$, and Hg$^{+}$. The calculations are used to find the
frequency shifts due to blackbody radiation which are needed for
accurate frequency measurements and improvements of the limits on
variation of the fine structure constant, $\alpha$. Our result for
Cs ($\delta \nu/E^2=-2.26(2)\times 10^{-10}$Hz/(V/m)$^2$) is in good
agreement with early measurements and {\em ab initio} calculations.
We present arguments against recent claims that the actual value
might be smaller. The difference ($\sim$ 10\%) is due to the
contribution of the continuum spectrum in the sum over intermediate
states.
\end{abstract}

 \maketitle

\section{Introduction}

Atomic clocks are now important for both practical applications and
fundamental physics. The hyperfine structure (hfs) transition of the
ground state of $^{133}$Cs serves as a primary frequency standard
providing the definition of a metric second. Many similar hfs
transitions in other atoms and ions are used or under consideration
for use as secondary microwave frequency standards. Most frequency
standards (atomic clocks) operate at room temperature. However, the
exact definition of a metric second corresponds to the frequency of
the transition measured at zero temperature. This means that
readings from atomic clocks should be corrected to account for the
effect of black body radiation (see e.g. \cite{Itano}). The value of
this effect can be found from measurements or calculations. There
are many experimental~\cite{Haun,Mowat,Bauch,Simon,Levi,Godone} and
theoretical~\cite{Itano,Feitchner,Lee,Palchikov,Micalizio,Ulzega,Beloy}
works studying the effects of black body radiation on microwave
frequency standards. However, the situation is far from being
satisfactory. There is disagreement among the different works for
cesium which will be discussed in more detail below. On the other
hand, the data for other atoms and ions is very poor or absent.

In the present work we perform accurate calculations of the shift
due to blackbody radiation in the hyperfine transitions in Yb$^{+}$,
Rb, Cs, Ba$^{+}$, and Hg$^{+}$. These transitions play key roles in
microwave frequency standards. They are also  of interest to
physicists considering experiments to measure $\alpha$ variation in
the laboratory (see e.g.  \cite{Peik, Karshemboim, Fischer, Bize}).
An experiment is currently planned utilizing Yb$^{+}$~\cite{Bruce}.
No accurate calculations or measurements of the radiation shift for
this ion have been performed previously. However, calculations or
measurements are available for other atoms and ions
\cite{Itano,Haun,Mowat,Bauch,Simon,Levi,Godone,Feitchner,Lee,Palchikov,Micalizio,Ulzega,Beloy}.
We compare our results with these values.

Due to the importance of cesium as a primary frequency standard we
have performed a more detailed study of this atom. This was the
subject of a separate short paper~\cite{Angstmann}. In the present
paper we give more details of the calculations while also discussing
other atoms and ions.

There is some disagreement on the actual value of the radiation
frequency shift for cesium. Early
measurements~\cite{Haun,Mowat,Simon} and {\em ab initio}
calculations~\cite{Lee,Palchikov} support a value which is close to
$-2.2\times 10^{-10}$Hz/(V/m)$^2$ while more recent
measurements~\cite{Levi,Godone} and semiempirical
calculations~\cite{Feitchner,Micalizio,Ulzega} claim that the actual
number might be about 10\% smaller. While we cannot comment on the
experimental results, the source of disagreement between theoretical
values seems to be in the continuum spectrum. {\em Ab initio}
calculations (including the present work) include the contribution
of the continuum spectrum into the summation over the complete set
of states while semiempirical calculations do not. We demonstrate
that adding the contribution of the continuum spectrum to where it
was missed brings all theoretical results into good agreement with
each other and with early measurements.

\section{Theory}

Blackbody radiation creates a temperature dependent electric field,
described by the Planck radiation law
\begin{equation}
E^{2}(\omega) = \frac{8 \alpha}{\pi}\frac{\omega ^{3} d
\omega}{\textrm{exp}(\omega/k_BT)-1}.  \label{Planck}
\end{equation}
This leads to the following expression for the average electric
field radiated by a black body at temperature T:
\begin{equation}\label{electric_field}
    \langle E^{2} \rangle = (831.9 \textrm{V/m})^{2}[\textrm{T(K)/300}]^{4}.
\end{equation}
This electric field causes a temperature-dependent frequency shift
of the atomic microwave clock transitions. It can be presented in
the form (see, e.g.~\cite{Itano})
\begin{equation}\label{beta}
    \delta \nu/\nu_0 = \beta (T/T_0)^4 \left[ 1 + \epsilon (T/T_0)^2
    \right] .
\end{equation}
Here $T_0$ is usually assumed to be room temperature ($T_0 = 300K$).
The frequency shift in a static electric field is
\begin{equation}\label{ke2}
    \delta \nu = k E^2.
\end{equation}
The coefficients $k$ and $\beta$ are related by
\begin{equation}
    \beta = \frac{k}{\nu_0} (831.9 \textrm{V/m})^2 \label{betak},
\end{equation}
$\epsilon$ is a small correction due to frequency distribution
(\ref{Planck}). In the present work we consider both terms in (\ref{beta}) by calculating coefficients $k$, $\beta$ and $\epsilon$.

It is convenient to start from calculation of $k$ by considering an atom in static electric field.
In the case when there is no other external electric field which sets preferred direction
the radiation shift can be expressed in terms of the scalar
hyperfine polarizability of the atom. This corresponds to averaging
over all possible directions of the electric field. The hyperfine
polarizability is the difference of the atomic polarizabilities
between different hyperfine structure states of the atom. The
lowest-order effect is linear in the hyperfine interaction and
quadratic in the electric field. Therefore, its value can be
calculated using third-order perturbation theory (see, e.g.
\cite{Landau})
\begin{eqnarray}
\delta \epsilon_a &=&  \sum_{n,m} \frac{\langle a | \hat V | n \rangle
\langle n | \hat V | m \rangle \langle m | \hat V | a \rangle}
{(\epsilon_a - \epsilon_n)(\epsilon_a - \epsilon_m)} \nonumber \\
 &-&  \langle a | \hat V | a \rangle
\sum_{n} \frac{\langle a | \hat V | n \rangle^2}
{(\epsilon_a - \epsilon_n)^2}.
\label{eq:3}
\end{eqnarray}
In our case the perturbation operator $\hat V$ is the sum of the
hyperfine structure operator and the electric dipole operator
\[ \hat V = \hat H_{hfs} -e\mbox{\boldmath{$E$}}\cdot\mbox{\boldmath{$r$}}.  \]
The operator of the hyperfine interaction $\hat H_{hfs}$ is given by
\begin{equation}\label{eq:hyperfine}
\hat{H}_{hfs} = \frac{|e|}{c}\mbox{\boldmath{$\mu$}} \cdot
\frac{\mbox{\boldmath{$r$}}\times\mbox{\boldmath{$\alpha$}}}
{r_>^{3}}, \ \ \ r_> =\max(r,r_N),
\end{equation}
where {\boldmath{$\alpha$}} is the Dirac matrix,
{\boldmath{$\mu$}} is the magnetic moment of the nucleus and $r_N$
is nuclear radius.

To get the effect of the electric field on the frequency of the
hyperfine transition one needs to go through the following steps:
\begin{itemize}
    \item Substitute the perturbation operator $\hat V$ into equation (\ref{eq:3}).
    \item Keep only terms linear in $\hat H_{hfs}$ and quadratic in the electric field.
    \item Apply equation (\ref{eq:3}) to both components of the hyperfine structure doublet.
    \item Take the difference.
\end{itemize}
The resulting expression for the frequency shift consists of three
terms. The first two of them originate from the first term of
equation (\ref{eq:3}). In one of them the $\hat H_{hfs}$ operator is
either on the left or right side of the expression (these two terms
are equal and can be combined together), and in the other the $\hat
H_{hfs}$ operator is in the middle. The last term is due to change
of the normalization of the wave function (second term of equation
(\ref{eq:3})). It is proportional to the hyperfine structure of the
ground state.

After angular reduction these three terms become
\begin{eqnarray}
\label{nu1}
 && \delta \nu_1(as) = e^2 \langle E^2 \rangle \frac{2I+1}{6} \times \nonumber \\
 && \sum_{n,m,j}
   \frac{ A_{as,ns} \langle ns || r || mp_j \rangle \langle mp_j || r || as \rangle}
  {(\epsilon_{as} - \epsilon_{ns})(\epsilon_{as} - \epsilon_{mp_j})},
\end{eqnarray}

\begin{eqnarray}
\label{nu2}
 && \delta \nu_2(as) = \frac{e^2 \langle E^2 \rangle }{6}
  \sum_{j}(C_{I+1/2} - C_{I-1/2}) \times \nonumber \\
 && \sum_{n,m} \frac{\langle as||r||npj \rangle
   A_{npj,mpj}  \langle mp_j || r || as \rangle}
  {(\epsilon_{as} - \epsilon_{npj})(\epsilon_{as} - \epsilon_{mp_j})},
\end{eqnarray}
and
\begin{eqnarray}
\label{norm}
 &&\delta \nu_{norm}(as) = \\
 &&- e^2 \langle E^2 \rangle \frac{2I+1}{12}
  A_{as} \sum_{m,j}
    \frac{|\langle as || r || mp_j \rangle|^2}
  {(\epsilon_{as} - \epsilon_{mp_j})^2}. \nonumber
\end{eqnarray}
Here
\begin{eqnarray}
&&C_F = \sum_{F'} (2F'+1)[F'(F'+1)-I(I+1)-j(j+1)] \nonumber \\
&& \times \left\{ \begin{array}{ccc} 1/2 & F & I \\ F' & j & 1
\end{array} \right\}^2, \ \ \ F'=|I-J|, I+J,
\nonumber
\end{eqnarray}
$A_{ns}$ is the hfs constant of the $ns$ state, $A_{m,n}$ is the
off-diagonal hfs matrix element, $I$ is nuclear spin,
$\mathbf{F=I+J}$, $\mathbf{J}$ is total electron momentum of the
atom in the ground state ($J=1/2$ for atoms considered in present
work), and $j$ is total momentum of virtual $p$-states
($j=1/2,3/2$). Summation goes over a complete set of $ns$,
$mp_{1/2}$ and $mp_{3/2}$ states.

Expression (\ref{nu1}) does not include the $s-d$ hfs matrix elements
while expression (\ref{nu2}) does not include the $p_{1/2}-p_{3/2}$
hfs matrix elements. Test calculations show that the total contribution 
of the off-diagonal (in total momentum~$j$) hfs matrix elements is of the order of 0.1\% of the
final answer and therefore can be neglected in present calculations.

Expressions (\ref{nu1}), (\ref{nu2}) and (\ref{norm}) correspond to static limit when energy of thermal photon is zero. To take into account distribution (\ref{Planck}) one needs to make the following substitutions in terms (\ref{nu1}) and (\ref{nu2}):
\begin{equation}
	\frac{\langle mp_j || r || as \rangle}{\Delta \epsilon_{sp}} \rightarrow
	\frac{1}{2}\left[\frac{\langle mp_j || r || as \rangle}{\Delta \epsilon_{sp} + \omega} +
	\frac{\langle mp_j || r || as \rangle}{\Delta \epsilon_{sp} - \omega}\right],
\end{equation}
and in term (\ref{norm})
\begin{equation}
	\frac{1}{\Delta \epsilon_{sp}^2} \rightarrow \frac{1}{2}\left[
	\frac{1}{(\Delta \epsilon_{sp} + \omega)^2} +
		\frac{1}{(\Delta \epsilon_{sp} - \omega)^2}\right],
\end{equation}
where $\omega$ is the frequency of thermal photon. Integrating resulting expression with (\ref{Planck}) and keeping only terms up to $\omega^2$ (since $\omega \ll \Delta \epsilon_{sp}$) leads to expression of the form (\ref{beta}) in which first term is given by (\ref{nu1}), (\ref{nu2}) and (\ref{norm}) and parameter $\epsilon$ in second term is given by
\begin{equation}
	\epsilon = \frac{A}{k} \left[\sum_i \frac{k_i^{(1)}}{\Delta \epsilon_{spi}^2} +
	\sum_i \frac{k_i^{(2)}}{\Delta \epsilon_{spi}^2} + 3\sum_i \frac{k_i^{(3)}}{\Delta \epsilon_{spi}^2}\right].
	\label{epsilon}
\end{equation}
Here index $i$ replaces all indexes of summation in (\ref{nu1}), (\ref{nu2}) and (\ref{norm}), $k_i^{(1)}$ corresponds to terms in (\ref{nu1}), $k_i^{(2)}$ corresponds to (\ref{nu2}), $k_i^{(3)}$ corresponds to (\ref{norm}) and $k = k^{(1)} +  k^{(2)} +  k^{(3)}$. $\Delta \epsilon_{spi}$ is the energy of the $s-p$ transition number $i$. If energies $\Delta \epsilon_{spi}$ are in atomic units then $A=1.697 \times 10^{-5}$
(atomic unit of energy is 27.211 eV = 315773K). Lowest $s -p $ transitions (e.g., $6s - 6p_{1/2}$ and $6s - 6p_{3/2}$) strongly dominate in the summation (\ref{epsilon}).

\section{Calculations}

In order to calculate frequency shift to the hfs transitions due to
the electric field one needs to have a complete set of states and to
have the energies, electric dipole transition amplitudes and
hyperfine structure matrix elements corresponding to these states.
It is possible to consider summation over the physical states and
use experimental data to perform the calculations. The lowest
valence states for which experimental data is usually available
dominate in the summation. Off-diagonal hfs matrix elements can be
obtained to a high accuracy as the square root of the product of
corresponding hfs constants: $A_{m,n} = \sqrt{A_m A_n}$ (see,
e.g.~\cite{offdhfs}). However, the accuracy of this approach is
limited by the need to include the {\it tail} contribution from
highly excited states including states in the continuum. This
contribution can be very significant and its calculation is not
easier than the calculation of the whole sum. Also, for atoms like
Yb$^+$ and Hg$^+$ available experimental data is insufficient to
follow this path.

Therefore, in the present work we use an {\em ab initio} approach in
which high accuracy is achieved by the inclusion of all important
many-body and relativistic effects. We make only one exception
toward the semiempirical approach. The frequency shift is dominated
by the term (\ref{norm}) which is proportional to the hfs in the
ground state. These hfs constants are known to very high accuracy
from measurements for all atoms considered in the present work. It
is natural to use experimental hfs constants in the dominating term
to have more accurate results. Note however that the difference with
complete {\it ab initio} calculations is small.
It is also instructive to perform
calculations of the hfs and atomic polarizabilities to demonstrate the
accuracy of the method.

Calculations start from the relativistic Hartree-Fock (RHF) method
in the $V^{N-1}$ approximation. This means that the initial RHF
procedure is done for a closed-shell atomic core with the valence
electron removed. After that, states of the external electron are
calculated in the field of the frozen core. Correlations are
included by means of the correlation potential method~\cite{CPM}. We
use two different approximations for the correlation potential,
$\hat \Sigma$. First, we calculate it in the lowest, second-order of
the many-body perturbation theory (MBPT). We use notation $\hat
\Sigma^{(2)}$ for the corresponding correlation potential. Then we
also include into $\hat \Sigma$ two classes of the higher-order
terms: screening of the Coulomb interaction and hole-particle
interaction (see, e.g.~\cite{all-order} for details). These two
effects are included in all orders of the MBPT and the corresponding
correlation potential is named $\hat \Sigma^{(\infty)}$.

To calculate $\hat \Sigma^{(2)}$ we need a complete set of
single-electron orbitals. We use the B-spline technique
\cite{Johnson1, Johnson3} to construct the basis. The orbitals are
built as linear combinations of 50 B-splines in a cavity of radius
40$a_B$. The coefficients are chosen from the condition that the
orbitals are eigenstates of the RHF Hamiltonian $\hat H_0$ of the
closed-shell core. The $\hat \Sigma^{(\infty)}$ operator is
calculated with the technique which combines solving equations for
the Green functions (for the direct diagram) with the summation over
complete set of states (exchange diagram)~\cite{all-order}.

The correlation potential $\hat \Sigma$ is then used to build a new
set of single-electron states, the so-called Brueckner orbitals.
This set is to be used in the summation in equations (\ref{nu1}),
(\ref{nu2}) and (\ref{norm}). Here again we use the B-spline
technique to build the basis. The procedure is very similar to the
construction of the RHF B-spline basis. The only difference is that
new orbitals are now the eigenstates of the $\hat H_0 + \hat \Sigma$
Hamiltonian, where $\hat \Sigma$ is either $\hat \Sigma^{(2)}$ or
$\hat \Sigma^{(\infty)}$.

We use the all-order correlation potential $\hat \Sigma^{(\infty)}$
for Rb, Cs and Ba$^+$. It has been demonstrated in a number of works
(see, e.g.~\cite{all-order,francium,s-d}) that inclusion of the
screening of Coulomb interaction and the hole-particle interaction
leads to very accurate results for alkali atoms. However, for other
atoms with one external electron above closed shells these two
higher-order effects are not dominating and their inclusion
generally does not improve the results. Therefore, for the Yb$^+$
and Hg$^+$ ions we use only the second-order correlation potential
$\hat \Sigma^{(2)}$.

\begin{table}
    \caption{Rescaling parameters for the $\hat \Sigma$ operator which fit energies of the lowest $s$ and $p$
    states of Rb, Cs, Ba,$^+$, Yb$^+$ and Hg$^+$.}
    \begin{ruledtabular}
        \begin{tabular}{l l l l l}
        \multicolumn{1}{c}{Atom} & \multicolumn{1}{c}{$\hat \Sigma$} & \multicolumn{1}{c}{$s_{1/2}$} &
        \multicolumn{1}{c}{$p_{1/2}$} & \multicolumn{1}{c}{$p_{3/2}$}  \\
        \hline
        Rb     & $\hat \Sigma^{(2)}$      &   0.868 &   0.903 &   0.906 \\
        Rb     & $\hat \Sigma^{(\infty)}$ &   1.008 &   0.974 &   0.976 \\
        Cs     & $\hat \Sigma^{(2)}$      &   0.802 &   0.848 &   0.852 \\
        Cs     & $\hat \Sigma^{(\infty)}$ &   0.985 &   0.95  &   0.95 \\
        Ba$^+$ & $\hat \Sigma^{(2)}$      &   0.782 &   0.830 &   0.833 \\
        Ba$^+$ & $\hat \Sigma^{(\infty)}$ &   0.988 &   0.963 &   0.967 \\
        Yb$^+$ & $\hat \Sigma^{(2)}$      &   0.866 &   1.09  &   1.185 \\
        Hg$^+$ & $\hat \Sigma^{(2)}$      &   0.805 &   0.890 &   0.926 \\
                \end{tabular}
        \label{fit}
        \end{ruledtabular}
\end{table}

Brueckner orbitals which correspond to the lowest valence states are
good approximations to the real physical states. Their quality can
be checked by comparing experimental and theoretical energies.
Moreover, their quality can be further improved by rescaling the
correlation potential $\hat \Sigma$ to fit experimental energies
exactly. We do this by replacing the $\hat H_0 + \hat \Sigma$ with
the $\hat H_0 + \lambda \hat \Sigma$ Hamiltonian in which the
rescaling parameter $\lambda$ is chosen for each partial wave to fit
the energy of the first valence state. The values of $\lambda$ are
presented in Table~\ref{fit}. Note that all values are very close to
unity. This means that even without rescaling the accuracy is very
good and only a small adjustment to the value of $\hat \Sigma$ is
needed. Note also that since the rescaling procedure affects both
energies and wave functions, it usually leads to improved values of
the matrix elements of external fields. In fact, this is a
semiempirical method to include omitted higher-order correlation
corrections.

Matrix elements of the hfs and electric dipole operators are found by means of the
time-dependent Hartree-Fock (TDHF) method~\cite{CPM,TDHF}. This method is equivalent
to the well-known random-phase approximation (RPA). In the TDHF method, single-electron
wave functions are presented in the form $\psi = \psi_0 + \delta \psi$, where $\psi_0$
is unperturbed wave function. It is an eigenstate of the RHF Hamiltonian $\hat H_0$:
$(\hat H_0 -\epsilon_0)\psi_0 = 0$.  $\delta \psi$ is the correction due to external
field. It can be found be solving the TDHF equation
\begin{equation}
    (\hat H_0 -\epsilon_0)\delta \psi = -\delta\epsilon \psi_0 - \hat F \psi_0 - \delta \hat V^{N-1} \psi_0,
    \label{TDHF}
\end{equation}
where $\delta\epsilon$ is the correction to the energy due to external field ($\delta\epsilon\equiv 0$
for the electric dipole operator), $\hat F$ is the operator of the external field
($\hat H_{hfs}$ or $e\mbox{\boldmath{$E$}}\cdot \mbox{\boldmath{$r$}}$), and $\delta \hat V^{N-1}$ is
the correction to the self-consistent potential of the core due to external field.
The TDHF equations are solved self-consistently for all states in the core. Then
matrix elements between any (core or valence) states $n$ and $m$ are given by
\begin{equation}
    \langle \psi_n | \hat F + \delta \hat V^{N-1} | \psi_m \rangle.
    \label{mel}
\end{equation}
The best results are achieved when $\psi_n$ and $\psi_m$ are Brueckner orbitals
calculated with rescaled correlation potential $\hat \Sigma$.

We use equation (\ref{mel}) for all hfs and electric dipole matrix
elements in (\ref{nu1}), (\ref{nu2}), and (\ref{norm}) except for
the ground state hfs matrix element in (\ref{norm}) where we use experimental data.

To check the accuracy of the calculations we perform calculations of
the hfs in the ground state and of the static scalar
polarizabilities of the atoms. Polarizabilities are given by the
expression
\begin{equation}
    \alpha_0(a) =\frac{2}{3} \sum_m \frac{|\langle a || r || m \rangle|^2}
  {(\epsilon_{a} - \epsilon_{m})}
  \label{alpha0}
\end{equation}
which is very similar to the term (\ref{norm}) for the frequency
shift. The most important difference is that the energy denominator
is squared in term (\ref{norm}) but not in (\ref{alpha0}). This
means better convergence with respect to the summation over complete
set of states for term (\ref{norm}) than for (\ref{alpha0}).
Therefore, if good accuracy is achieved for polarizabilities, even
better accuracy should be expected for the term (\ref{norm}) (see
also Ref.~\cite{Micalizio}).

However, the behavior of the other two terms, (\ref{nu1}) and
(\ref{nu2}), is very different and calculation of polarizabilities
tells us little about accuracy for these terms.
Therefore, we also perform detailed calculations of the hfs
constants of the ground state. Inclusion of core polarization
(second term in (\ref{mel})) involves summation over the complete
set of states similar to what is needed for term (\ref{nu1}).
Comparing experimental and theoretical hfs is a good test of the
accuracy of this term.

\begin{table}
    \caption{Contributions to the hyperfine structure of the ground state of Rb, Cs, Ba,$^+$, Yb$^+$ and Hg$^+$ (MHz);
    comparison with experiment.}
    \begin{ruledtabular}
        \begin{tabular}{l l r r r r r r}
        \multicolumn{2}{c}{Atom} & \multicolumn{1}{c}{Brueck} & \multicolumn{1}{c}{RPA} & \multicolumn{1}{c}{Str} &
        \multicolumn{1}{c}{Norm} & \multicolumn{1}{c}{Total} & \multicolumn{1}{c}{Exp} \\
        \hline
        $^{87}$Rb      & $5s$ &   2888 &    559 &   -27 &   -33 &   3386  &   3417\footnotemark[1] \\
        $^{133}$Cs     & $6s$ &   1957 &    355 &   -10 &   -31 &   2278  &   2298\footnotemark[2] \\
        $^{137}$Ba$^+$ & $6s$ &   3509 &    601 &   -21 &   -73 &   4016  &   4019\footnotemark[3] \\
        $^{171}$Yb$^+$ & $6s$ &  11720 &   1540 &  -248 &  -247 &  12764  & 12645(2)\footnotemark[4] \\
        $^{199}$Hg$^+$ & $6s$ &  38490 &   3873 &   288 & -1483 &  41169  & 40507\footnotemark[5] \\
        \end{tabular}
    \footnotetext[1]{Reference \cite{CsRb_hfs}.}
    \footnotetext[2]{Reference \cite{CsRb_hfs}.}
    \footnotetext[3]{References  \cite{Ba_hfs1} and \cite{Ba_hfs2}.}
    \footnotetext[4]{Reference \cite{Yb_hfs}.}
    \footnotetext[5]{Reference \cite{Hg_hfs}.}
    \label{hfs}
        \end{ruledtabular}
\end{table}

In addition to term (\ref{mel}), we also include two smaller
contributions to the hfs: structure radiation and the correction due
to the change of the normalization of the wave function. The
structure radiation term can be presented in the form
\begin{equation}
    A_{str} = \langle \psi_n | \delta \hat \Sigma | \psi_n \rangle,
    \label{str}
\end{equation}
where $\delta \hat \Sigma$ is the correction to the correlation
potential due to external hfs field. The normalization term is
\begin{equation}
    A_{norm} = - A_n \langle \psi_n | \frac{\partial \hat \Sigma}{\partial\epsilon} | \psi_n \rangle,
    \label{norma}
\end{equation}
where $A_n$ is given by (\ref{mel}) with $m=n$.

The results for hfs are presented in Table~\ref{hfs}. Here column
marked as 'Brueck' corresponds to the $\langle \psi_n | \hat F |
\psi_n \rangle$ matrix element. The column marked as RPA is the core
polarization correction given by $\langle \psi_n |\delta \hat
V^{N-1} | \psi_n \rangle$, the notation 'Str' stands for structure
radiation given by (\ref{str}), and 'Norm' is the renormalization
contribution given by (\ref{norma}). In all cases $\psi_n$ is the
Bruckner orbital corresponding to the ground state of the atom or
ion, calculated with the rescaled correlation potential $\hat
\Sigma$. All-order $\hat \Sigma^{(\infty)}$ is used for Rb, Cs and
Ba$^+$. Second-order $\hat \Sigma^{(2)}$ is used for Yb$^+$ and
Hg$^+$.  Comparing the final theoretical results with experiment
shows that the theoretical accuracy is within 1\% for all atoms
except Hg$^+$ where it is 1.6\%. If the structure radiation and
normalization are neglected, accuracy for Rb and Cs remains within
1\%, accuracy for Ba$^+$ becomes about 2\% and accuracy for Yb$^+$
and Hg$^+$ becomes close to 5\%.

The results for polarizabilities, calculated in different
approximations, are presented in Table~\ref{pol0}. Pure {\it ab
initio} results obtained with $\hat \Sigma^{(\infty)}$ and results
obtained with rescaled correlation potential operators $\hat
\Sigma^{(2)}$ and $\hat \Sigma^{(\infty)}$ are very close to each
other and to other calculations and measurements.

\begin{table}
    \caption{Static polarizabilities $\alpha_0$ of Rb, Cs, Ba,$^+$, Yb$^+$ and Hg$^+$ in different approximations ($a_0^3$).}
    \begin{ruledtabular}
        \begin{tabular}{l l l r r r r r}
        \multicolumn{2}{c}{Atom} & \multicolumn{1}{c}{$\hat \Sigma$} & \multicolumn{1}{c}{$\alpha_v$\footnotemark[1]} & \multicolumn{1}{c}{$\alpha_c$\footnotemark[2]} &
        \multicolumn{1}{c}{Total}  & \multicolumn{1}{c}{Other works} \\
        \hline
        $^{87}$Rb      & $5s$ & $\hat \Sigma^{(2)}$\footnotemark[3]
                                                  & 292.7 &  9.1 & 301.8 &   329(23)\footnotemark[6] \\
                       &      & $\lambda \hat \Sigma^{(2)}$\footnotemark[4]
                                                  & 309.7 &  9.1 & 318.8 &   293(6)\footnotemark[7]    \\
                       &      & $\hat \Sigma^{(\infty)}$\footnotemark[5]
                                                  & 312.4 &  9.1 & 321.5 &  318.6(6)\footnotemark[8] \\
                       &      & $\lambda \hat \Sigma^{(\infty)}$\footnotemark[4]
                                                  & 310.5 &  9.2 & 319.7 & 318.5(6)\footnotemark[9] \\

        $^{133}$Cs     & $6s$ & $\hat \Sigma^{(2)}$\footnotemark[3]
                                              & 343.8 & 15.3 & 359.1 & 401.0(6)\footnotemark[10] \\
                       &      & $\lambda \hat \Sigma^{(2)}$\footnotemark[4]
                                              & 383.5 & 15.4 & 399.0 &  401.5\footnotemark[8]     \\
                       &      & $\hat \Sigma^{(\infty)}$\footnotemark[5]
                                              & 384.0 & 15.5 & 399.5 & 400.4\footnotemark[11] \\
                       &      & $\lambda \hat \Sigma^{(\infty)}$\footnotemark[4]
                                              & 384.4 & 15.5 & 399.9 & 400.6(1.0)\footnotemark[12] \\

        $^{137}$Ba$^+$ & $6s$ & $\hat \Sigma^{(2)}$\footnotemark[3]
                                               & 104.1 &  9.8 & 113.8 &         \\
                       &      & $\lambda \hat \Sigma^{(2)}$\footnotemark[4]
                                               & 112.5 &  9.9 & 122.4 &    \\
                       &      & $\hat \Sigma^{(\infty)}$\footnotemark[5]
                                               & 112.8 &  9.9 & 122.7 &    \\
                       &      & $\lambda \hat \Sigma^{(\infty)}$\footnotemark[4]
                                               & 112.7 &  9.9 & 122.7 &    \\

        $^{171}$Yb$^+$ & $6s$ & $\hat \Sigma^{(2)}$\footnotemark[3]
                                               &  50.9 &  6.2 &  57.1 &          \\
                       &      & $\lambda \hat \Sigma^{(2)}$\footnotemark[4]
                                               &  55.4 &  6.1 &  61.5 &    \\

        $^{199}$Hg$^+$ & $6s$ & $\hat \Sigma^{(2)}$\footnotemark[3]
                                               &  10.5 &  7.7 &  18.2 &          \\
                       &      & $\lambda \hat \Sigma^{(2)}$\footnotemark[4]
                                               &  11.4 &  7.6 &  19.0 &    \\
        \end{tabular}
    \footnotetext[1]{Polarizability due to valence electron.}
    \footnotetext[2]{Polarizability of the core.}
    \footnotetext[3]{$\hat \Sigma^{(2)}$ is the second-order correlation potential.}
    \footnotetext[4]{Rescaled $\hat \Sigma$. See Table~\ref{fit} for the values of rescaling factors $\lambda$.}
    \footnotetext[5]{$\hat \Sigma^{(\infty)}$ is the all-order correlation potential.}
      \footnotetext[6]{Measurements, Ref.~\cite{Molof}.}
      \footnotetext[7]{Measurements, Ref.~\cite{Hall}.}
      \footnotetext[8]{Calculations, Ref.~\cite{Derevianko}.}
      \footnotetext[9]{Calculations, Ref.~\cite{Safronova}.}
      \footnotetext[10]{Measurements, Ref.~\cite{Amini}.}
      \footnotetext[11]{Calculations, Ref.~\cite{Porsev}.}
      \footnotetext[12]{Calculations, Ref.~\cite{Micalizio}.}
    \label{pol0}
        \end{ruledtabular}
\end{table}

\section{Results and discussion}

Table \ref{terms} presents contributions of terms (\ref{nu1}),
(\ref{nu2}) and (\ref{norm}) into the total frequency shift of the
hfs transitions for the ground states of $^{87}$Rb, $^{133}$Cs,
$^{137}$Ba$^{+}$, $^{171}$Yb$^{+}$ and $^{199}$Hg$^{+}$ calculated
in different approximations. Term (\ref{norm}) dominates in all
cases, while term (\ref{nu2}) is small but still important at least
for Rb, Cs and Ba$^+$. Results obtained with $\hat \Sigma^{(2)}$ and
$\hat \Sigma^{(\infty)}$ differ significantly (up to 14\% for Cs).
However, after rescaling the results for both $\hat \Sigma^{(2)}$
and $\hat \Sigma^{(\infty)}$ come within a fraction of a per cent of
each other. Naturally, rescaling has a larger effect on results
obtained with $\hat \Sigma^{(2)}$. This means that the rescaling
really imitates the effect of higher-order correlations and should
lead to more accurate results. Comparing the results obtained with
$\hat \Sigma^{(\infty)}$, rescaled $\hat \Sigma^{(\infty)}$ and
rescaled $\hat \Sigma^{(2)}$ gives a reasonable estimation of the
accuracy of calculations. If we also combine this with the
calculation of the hfs discussed above we can safely assume that the
accuracy of the calculations for Rb, Cs and Ba$^+$ is on the level
of 1\%. Note that the frequency shift due to black body radiation
can be a little larger ($\sim$~1\%) due to the effect of frequency
distribution at finite temperature.

For Yb$^+$ and Hg$^+$ we only have results with rescaled $\hat
\Sigma^{(2)}$ and not rescaled $\hat \Sigma^{(\infty)}$. They differ
by about 11\%. However, there are strong reasons to believe that the
results obtained with rescaling are more accurate. This is supported
by calculations for Rb, Cs and Ba$^+$ as well as our experience with
rescaling used in many earlier works. Therefore, the calculation of
the hfs discussed above gives a more realistic estimation of the
accuracy for Yb$^+$ and Hg$^+$ which is about 5\%.

\begin{table}
    \caption{Contribution of terms (\ref{nu1}), (\ref{nu2}), and (\ref{norm}) to the frequencies of the hyperfine
    transitions in the ground state of Rb, Cs, Ba,$^+$, Yb$^+$ and Hg$^+$ ($\delta\nu_{0}$/$E^{2} \times 10^{-10}$ Hz/(V/m)$^2$) in different approximations.}
    \begin{ruledtabular}
        \begin{tabular}{l l l r r r r r}
        \multicolumn{2}{c}{Atom} & \multicolumn{1}{c}{$\hat \Sigma$} & \multicolumn{1}{c}{(\ref{nu1})} & \multicolumn{1}{c}{(\ref{nu2})} &
        \multicolumn{1}{c}{(\ref{norm})} & \multicolumn{1}{c}{Total}  \\
        \hline
        $^{87}$Rb      & $5s$ & $\hat \Sigma^{(2)}$\footnotemark[1]              & -0.5457 & 0.0147 & -0.6692 & -1.2003 \\
                       &      & $\lambda \hat \Sigma^{(2)}$\footnotemark[2]      & -0.5668 & 0.0154 & -0.6894 & -1.2409 \\
                       &      & $\hat \Sigma^{(\infty)}$\footnotemark[3]         & -0.5640 & 0.0156 & -0.7034 & -1.2518 \\
                       &      & $\lambda \hat \Sigma^{(\infty)}$\footnotemark[2] & -0.5620 & 0.0154 & -0.6972 & -1.2437 \\

        $^{133}$Cs     & $6s$ & $\hat \Sigma^{(2)}$\footnotemark[1]              & -0.9419 & 0.0210 & -1.0722 & -1.9931 \\
                       &      & $\lambda \hat \Sigma^{(2)}$\footnotemark[2]      & -1.0239 & 0.0229 & -1.2688 & -2.2697 \\
                       &      & $\hat \Sigma^{(\infty)}$\footnotemark[3]         & -1.0148 & 0.0232 & -1.2706 & -2.2622 \\
                       &      & $\lambda \hat \Sigma^{(\infty)}$\footnotemark[2] & -1.0167 & 0.0230 & -1.2695 & -2.2632 \\

        $^{137}$Ba$^+$ & $6s$ & $\hat \Sigma^{(2)}$\footnotemark[1]              & -0.1027 & 0.0034 & -0.1568 & -0.2561 \\
                       &      & $\lambda \hat \Sigma^{(2)}$\footnotemark[2]      & -0.1095 & 0.0036 & -0.1768 & -0.2827 \\
                       &      & $\hat \Sigma^{(\infty)}$\footnotemark[3]         & -0.1104 & 0.0037 & -0.1778 & -0.2845 \\
                       &      & $\lambda \hat \Sigma^{(\infty)}$\footnotemark[2] & -0.1104 & 0.0037 & -0.1773 & -0.2841 \\

        $^{171}$Yb$^+$ & $6s$ & $\hat \Sigma^{(2)}$\footnotemark[1]              & -0.0672 & 0.0009 & -0.0866 & -0.1529 \\
                       &      & $\lambda \hat \Sigma^{(2)}$\footnotemark[2]      & -0.0714 & 0.0011 & -0.1003 & -0.1706 \\

        $^{199}$Hg$^+$ & $6s$ & $\hat \Sigma^{(2)}$\footnotemark[1]              & -0.0242 & 0.0000 & -0.0296 & -0.0538 \\
                       &      & $\lambda \hat \Sigma^{(2)}$\footnotemark[2]      & -0.0263 & 0.0000 & -0.0335 & -0.0598 \\
        \end{tabular}
    \footnotetext[1]{$\hat \Sigma^{(2)}$ is the second-order correlation potential.}
    \footnotetext[2]{Rescaled $\hat \Sigma$. See Table \ref{fit} for the values of rescaling factors $\lambda$.}
    \footnotetext[3]{$\hat \Sigma^{(\infty)}$ is the all-order correlation potential.}

    \label{terms}
        \end{ruledtabular}
\end{table}

Table~\ref{terms} presents values of $k$ (see formula (\ref{ke2})).
To obtain the frequency shift at finite temperature one needs to
convert $k$ into $\beta$ using formula (\ref{betak}) and substitute
$\beta$ into equation (\ref{beta}). For accurate results one also
needs to know the values of $\epsilon$. We calculate them using formula (\ref{epsilon}) in a very similar way as we calculate parameters $k$. 
Our final values of $k$, $\beta$ and $\epsilon$ are presented in
Table~\ref{kbep}. Parameter $\epsilon$ for Cs was estimated in single-resonance approximation in Ref~\cite{Itano} and found to be 0.014. This value is in good agreement with our accurate calculations.

\begin{table}
\caption{Final results for the parameters $k$ ($10^{-10}$~Hz/(V/m)$^2$), $\beta$ ($10^{-14}$) and $\epsilon$ of the black-body radiation frequency shift for Rb, Cs, Ba,$^+$, Yb$^+$ and Hg$^+$.}
\begin{ruledtabular}
\begin{tabular}{l l l l l}
\multicolumn{2}{c}{Atom} & \multicolumn{1}{c}{$k$} & \multicolumn{1}{c}{$\beta$} & \multicolumn{1}{c}{$\epsilon$} \\
\hline
        $^{87}$Rb      & $5s$ & -1.24(1) & -1.26(1) & 0.011 \\
        $^{133}$Cs     & $6s$ & -2.26(2) & -1.70(2) & 0.013 \\
        $^{137}$Ba$^+$ & $6s$ & -0.284(3) & -0.245(2) & 0.004 \\
        $^{171}$Yb$^+$ & $6s$ & -0.171(9) & -0.094(5) & 0.002 \\
        $^{199}$Hg$^+$ & $6s$ & -0.060(3) & -0.0102(5) & 0.0005 \\
        \end{tabular}
    \label{kbep}
        \end{ruledtabular}
\end{table}

The frequency shifts of some alkali metal atoms have been calculated
and measured previously. We present previous results for the atoms
and ions for which we perform calculations in Table~\ref{results} together with our final results.

There is some disagreement for cesium. Our result is in good
agreement with early measurements~\cite{Haun,Mowat,Simon} and {\em
ab initio} calculations~\cite{Lee,Palchikov} while recent
measurements~\cite{Levi,Godone} and semiempirical
calculations~\cite{Feitchner,Micalizio,Ulzega} give the value which
is about 10\% smaller. Less accurate measurements of Bauch and
Schr\"{o}der~\cite{Bauch} cover both cases. We cannot comment on
disagreement between experimental results. However, the source of
disagreement between theoretical results seems to be clear. It comes
from the contribution of the continuum spectrum to the summation
over the complete set of states in term (\ref{nu1}). This term has
off-diagonal hfs matrix elements between the ground state and
excited states. Since the hfs interaction is localized over short
distances ($ \sim a_0/Z$) it emphasizes the contribution of states
with high energies including states in the continuum (since $\Delta
p \Delta x \geq\hbar$, a small area of localization ($\Delta x$)
allows high momentum ($p$) and thus high energy). In our
calculations the contribution of states above $7p$ into term
(\ref{nu1}) is $-0.35 \times 10^{-1}$Hz/(V/m)$^2$ which is 15\% of
the total answer.

In contrast, states above $7p$ contribute only about 0.05\% of the
total value of term (\ref{norm}). This is because the summation goes
over the matrix elements of the electric dipole operator which is
large on large distances and thus suppresses the contribution of
high-energy states. It is not surprising therefore that a
semiempirical consideration, involving only discrete spectrum
states, gives very good results for the atomic polarizabilities
(see, e.g.~\cite{Micalizio}). However, let us stress once more that
the calculation of polarizabilities checks only term (\ref{norm})
and tells us very little about the accuracy of other two terms,
(\ref{nu1}) and (\ref{nu2}).

The contribution of the states above $7p$ is even more important for
term (\ref{nu2}). Their contribution is about 30\% of the total
value of this term. However, the term itself is small and its
accurate treatment is less important.

In {\it ab initio} calculations by Lee {\it et al}~\cite{Lee}
summation over complete set of states is reduced to solving of a
radial equation (equations of this type are often called Sternheimer
equation after one of the authors of this work). This approach does
include the contribution of the continuum spectrum and the result is
in very good agreement with ours (see Table~\ref{results}).

In other {\it ab initio} calculations by Pal'chikov {\it et
al}~\cite{Palchikov} summation is done via Green functions. This
corresponds to summation over the complete set of states and does
include the continuum spectrum. Again, the result is in very good
agreement with other {\it ab initio} calculations (\cite{Lee} and
the present work).

Recent calculations by Beloy {\em et al}~\cite{Beloy} applied a mixed
approach, with extensive use of experimental data for lower cesium states
and {\em ab initio} summation over higher states including continuum.
The result is in good agreement with fully {\em ab initio} calculations.

In contrast, analysis performed in \cite{Feitchner,Micalizio,Ulzega}
is limited to discrete spectrum. Adding $-0.34 \times
10^{-1}$Hz/(V/m)$^2$ (which is total {\it tail} contribution from
all three terms (\ref{nu1}), (\ref{nu2}) and (\ref{norm}) found in
our calculation) to the results of Feitchner {\it
et~al}~\cite{Feitchner} and Micalizio {\it et~al}~\cite{Micalizio}
brings them to excellent agreement with {\it ab initio}
calculations. The same modification of the result by Ulzega {\it et
al}~\cite{Ulzega} makes it a little bit too large but still closer
to other results than without the {\it tail} contribution.

\begin{table}
    \caption{Electrostatic frequency shifts for the hyperfine
    transitions of Rb, Cs, Ba,$^+$, Yb$^+$ and Hg$^+$ ($\delta\nu_{0}$/$E^{2} \times 10^{-10}$ Hz/(V/m)$^2$) ; comparison with other calculations and measurements.}
    \begin{ruledtabular}
        \begin{tabular}{l l l l l l l}
        \multicolumn{2}{c}{Atom} & \multicolumn{1}{c}{This} & \multicolumn{1}{c}{Other} & \multicolumn{1}{c}{Ref} &    \multicolumn{1}{c}{Measurements} & \multicolumn{1}{c}{Ref} \\
                 &            & \multicolumn{1}{c}{work} & \multicolumn{1}{c}{calculations} &        \\
        \hline
        $^{87}$Rb      & $5s$ & -1.24(1)   & -1.2094 & \cite{Lee}       & -1.23(3)  & \cite{Mowat} \\

        $^{133}$Cs     & $6s$ & -2.26(2)   & -1.9(2)  & \cite{Feitchner} & -2.29(7)  & \cite{Haun}  \\
                       &      &         & -2.2302  & \cite{Lee}       & -2.25(5)  & \cite{Mowat} \\
                       &      &         & -2.28    & \cite{Palchikov} & -2.17(26) & \cite{Bauch} \\
                       &      &         & -1.97(9) & \cite{Micalizio} & -2.271(4) & \cite{Simon} \\
                       &      &         & -2.06(1) & \cite{Ulzega}    & -1.89(12) & \cite{Levi} \\
                       &      &         & -2.268(8)& \cite{Beloy}     & -2.03(4)  & \cite{Godone} \\

        $^{137}$Ba$^+$ & $6s$ & -0.284(3)  & -0.27   & \cite{Itano}     &           &              \\
        $^{171}$Yb$^+$ & $6s$ & -0.171(9)  &         &                  &           &              \\
        $^{199}$Hg$^+$ & $6s$ & -0.060(3) & -0.058  & \cite{Itano}     &           &              \\
                    \end{tabular}
    \label{results}
        \end{ruledtabular}
\end{table}

\section{Conclusion}

We have performed calculations of the frequency shift of the ground
state hyperfine transition for several atoms and ions caused by a
static electric field which can be used to evaluate the effect of
blackbody radiation on the frequency of the microwave atomic clock
transitions. The size of this shift is comparable to the current
error in the measurements of the energy shift caused by variation of
$\alpha$ and so needs to be taken into account in laboratory
measurements placing limits upon $\alpha$ variation.

Detailed analysis of the calculations for cesium reveal the source
of disagreement between different theoretical approaches. This seems
to be contribution of the continuum spectrum into summation over
complete set of states which was neglected in semiempirical
calculations. Restoring the {\it tail} contribution in works where
it was neglected brings all theoretical results in good agreement
with each other.

\section*{Acknowledgments}

This work was suggested by Bruce Warrington.
We are grateful for his initial suggestion and useful discussions.
We are also grateful to S. Ulzega, W. Itano and A. Derevianko for useful comments and references.
The work was supported by the Australian Research Council.

\end{document}